\documentclass[epj,nopacs]{svjour}
\usepackage{latexsym}
\usepackage{graphics}
\usepackage{graphicx}
\usepackage{bm}
\usepackage{epsfig}
\usepackage{amsmath,amssymb,amsfonts}
\usepackage{url}
\usepackage{epstopdf}
\usepackage{color}

\begin{document}

\title{Photon emission of extremal Kerr-Newman black holes}
\author{Shao-Wen Wei\thanks{\emph{e-mail:} weishw@lzu.edu.cn}
\and Bao-Min Gu\thanks{\emph{e-mail:} gubm15@lzu.edu.cn}
\and Yong-Qiang Wang\thanks{\emph{e-mail:} yqwang@lzu.edu.cn}
\and Yu-Xiao Liu\thanks{\emph{e-mail:} liuyx@lzu.edu.cn, corresponding author}}
\institute{Institute of Theoretical Physics, Lanzhou University, Lanzhou 730000, People's Republic of China}
\date{Received: date / Revised version: date}

\date{Received: date / Revised version: date}

\abstract{In this paper, we deal with the null geodesics extending from the near-horizon region out to a distant observatory in an extremal Kerr-Newman black hole background. In particular, using the matched asymptotic expansion method, we analytically solve the null geodesics near the superradiant bound in the form of algebraic equations. {For the case that the photon trajectories are limited in the equatorial plane, the shifts in the azimuthal angle and time are obtained.}}

%\keywords{Black holes, geodesic}

%\pacs{04.70.Dy, 04.25.-g}

\maketitle

\section{Introduction}
\label{secIntroduction}

The geodesic motion of test particles in a black hole background is one of the most interesting subjects in general relativity. Many astronomical phenomena are related with the geodesic, such as light deflection, perihelion shift of planets, Shapiro effect, Lense-Thirring, black hole shadow and so on. In the near future, the LIGO \cite{LIGO}, the Event Horizon Telescope \cite{EHT}, and other experiments \cite{ATHENA,SKA,eLISA} may have a qualitatively new level of precision to explore phenomena near the black hole horizon.

The Kerr-Newman (KN) black hole solution describes the gravitational field of a charged and rotating stationary black hole. Although it is not very likely that the astrophysical black hole candidates carry a net charge \cite{Eddington}, such charged black holes can be formed through a charged stellar collapse \cite{Ghezzi,Letelier}, brane-world-inspired charge-leaking mechanism \cite{Cuesta}, and accretion scenarios \cite{Damour,Ruffini}. The geodesic motion in this black hole background has been extensively investigated. For examples, the timelike equatorial and spherical orbits of uncharged particles and the last stable orbit of charged particles were studied in Refs. \cite{Johnston,Young}, respectively. The effects of back-reaction and observer-independence of speed of light for the fast spinning particle was considered in \cite{Deriglazov}. The motion of charged particles was also investigated in Ref. \cite{Bicak}, while the off-equatorial circular orbits were given in Ref. \cite{Kovar}. A comprehensive analysis of the photon orbit and charged particle motion in a KN black hole were presented in Refs. \cite{Calvani,Hackmann}. With the help of the multivariable hypergeometric functions of Appell-Lauricella and Weierstrass elliptic functions, the null geodesics of KN black hole are solved in a close form \cite{Kraniotis}. However, expressing the geodesics with simple functions is still an interesting issue.

Another theoretical interest for the KN black hole is the KN/CFTs (Conformal Field Theory) dualities \cite{Hartman,Song,Rasmussen,Chen,ChenChen,Sun,Bai}, which are the simple extension of the Kerr/CFT duality \cite{Guica,Bredberg,Compere}. It was also shown that the near-horizon geometry of a near extremal KN black hole has a warped $AdS_{3}=AdS_{2}\times S^{1}$ structure, and analytical results for the pair production were also obtained in Ref. \cite{Chiang}.

Very recently, Porfyriadis, Shi, and Strominger \cite{Porfyriadis} considered the photon emission near extremal Kerr black holes. And they first solved the near-superradiant geodesics analytically. Such results are hoped to apply to a variety of problems related to the observations of electromagnetic radiation. In this paper, we would like to generalize the result to the extremal KN black hole spacetime. First, we analyze the effective potential of the radial motion. We find a critical point, beyond which the photon emitting near the black hole horizon can escape the black hole and can be observed by distant observatories. Then near this critical point, we perform the geodesic integrals for both near and far regions. Finally, matching the solutions in the two regions, we obtain the analytic result for the extremal KN black hole. Using the result, observations in a KN spacetime are expected to be analytically researched.

Our paper is organized as follows. In the next section, we would like to give a brief review of the geodesic equations in the KN black hole background and analyze the effective potential. In Sec. \ref{rth} the null geodesic motion in the $r$-$\theta$ plane is analytically solved. And in Sec. \ref{rtrphi}, we deal with the motions in the $r$-$\phi$ and $r$-$t$ planes. At last, the paper ends with a brief summary.

\section{Geodesic equations and effective potential}

In this section, we would like to consider the geodesic equations in the background of a KN black hole.

\begin{figure}
\center{\includegraphics[width=6.5cm]{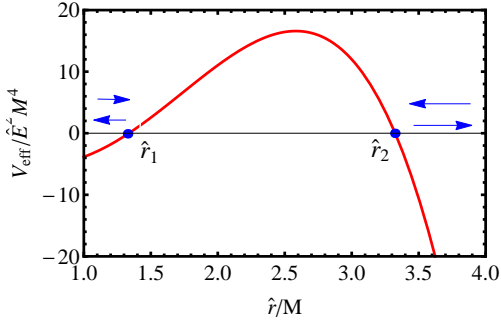}}
\caption{Effective potential $V_{eff}$ as a function of $r$ with $e=0.2$, $\hat{L}/M\hat{E}=4$ and $Q/M^{2}\hat{E}^{2}$=3.}\label{rhat1}
\end{figure}

The line element of a KN black hole in Boyer-Lindquist coordinates $(\hat{t}, \hat{r}, \theta, \hat{\phi})$ is
\begin{eqnarray}
 ds^{2}&=&-\frac{\Delta}{\rho^{2}}\left(d\hat{t}-a \sin^{2}\theta d\hat{\phi}\right)^{2}
        +\frac{\rho^{2}}{\Delta}d\hat{r}^{2}+\rho^{2}d\theta^{2}\nonumber\\
        &&+\frac{\sin^{2}\theta}{\rho^{2}}\left(ad\hat{t}-(\hat{r}^{2}+a^{2})d\hat{\phi}\right)^{2},
\end{eqnarray}
where the metric functions are
\begin{eqnarray}
 \Delta=\hat{r}^{2}-2M\hat{r}+a^{2}+\hat{e}^{2},\quad
 \rho^{2}=\hat{r}^{2}+a^{2}\cos^{2}\theta.
\end{eqnarray}
The vector potential is given by
\begin{eqnarray}
   A_{a}&=&\frac{\hat{e}\hat{r}}{\rho^{2}}(d\hat{t}-a\sin^{2}\theta d\hat{\phi}).
\end{eqnarray}
The parameter $a$, $\hat{e}$, and $M$ are the angular momentum, charge, and mass of the black hole. This black hole solution originates from the four-dimensional Einstein-Maxwell action
\begin{eqnarray}
 S=\frac{1}{16\pi}\int d^{4}x\sqrt{-g}(R-F^{2}).
\end{eqnarray}
The horizons are given by solving the equation $\Delta=0$, i.e.,
\begin{eqnarray}
 \hat{r}_{\pm}=M\pm\sqrt{M^{2}-a^{2}-\hat{e}^{2}}.
\end{eqnarray}
The extremal black hole is achieved when the two horizons coincide with each other. The motion of an electrically {neutral particle} moving in the KN black hole background is described by the Lagrangian
\begin{eqnarray}
 \mathcal{L}=-\frac{1}{2}g_{\mu\nu}\dot{x}^{\mu}\dot{x}^{\nu},
\end{eqnarray}
where a dot over a symbol represents the ordinary differentiation with respect to an affine parameter $\lambda$. The affine parameter is related to the proper time by $\tau=\mu\lambda$, which is equivalent to the normalization condition  $g_{\mu\nu}\dot{x}^{\mu}\dot{x}^{\nu}=-\mu^{2}$. The equation of motion is 
\begin{eqnarray}
 \frac{d^{2}x^{\sigma}}{d\tau^{2}}
      +\Gamma^{\sigma}_{\;\mu\nu}\frac{dx^{\mu}}{d\tau}\frac{dx^{\nu}}{d\tau}
      =0,
\end{eqnarray}
where $\Gamma^{\sigma}_{\;\mu\nu}$ is the Christoffel symbols for the KN black hole. Solving it, the equation of motion will be \cite{Carter}
\begin{eqnarray}
 &&\int^{\hat{r}}\frac{d\hat{r}'}{\sqrt{\hat{R}}}=\int^{\theta}\frac{d\theta'}{\sqrt{\hat{\Theta}}},\label{ttheta}\\
 &&\hat{\phi}=\int^{\hat{r}}\frac{a\hat{E}\hat{r}'^{2}+(\hat{L}-a\hat{E})(\Delta-a^{2})}{\Delta\sqrt{\hat{R}}}d\hat{r}'
            \nonumber\\
            &&\quad\quad+\int^{\theta}\frac{\hat{L}\cot^{2}\theta'}{\sqrt{\hat{\Theta}}}d\theta',\\
 &&\hat{t}=\int^{\hat{r}}\frac{\hat{E}\hat{r}'^{2}(\hat{r}'^{2}+a^{2})+a(\hat{L}-a\hat{E})(\Delta-\hat{r}'^{2}-a^{2})}
         {\Delta\sqrt{\hat{R}}}d\hat{r}'\nonumber\\
         &&\quad\quad+\int^{\theta}\frac{a^{2}\hat{E}\cos^{2}\theta'}{\sqrt{\hat{\Theta}}}d\theta'.\label{ttheta2}
\end{eqnarray}
with
\begin{eqnarray}
 \hat{R}&=&\left(\hat{E}(\hat{r}^{2}+a^{2})-\hat{L}a\right)^{2}\nonumber\\
                 &&-\Delta(\mu^{2}\hat{r}^{2}+(\hat{L}-a\hat{E})^{2}+Q),\\
 \hat{\Theta}&=&Q-\cos^{2}\theta\left(a^{2}(\mu^{2}-\hat{E}^{2})+\hat{L}^{2}/\sin^{2}\theta\right),
\end{eqnarray}
where $\hat{E}$, $\hat{L}$, and $Q$ are the energy, angular momentum, and Carter constant, respectively. When the black hole charge $\hat{e}=0$, the result for the Kerr black hole will be recovered. Next, we will consider the null geodesics in an extremal KN black hole background, i.e.,
\begin{eqnarray}
 \mu^{2}=0,\quad a^{2}=M^{2}(1-e^{2}).
\end{eqnarray}
Here $e=\hat{e}/M$ is the dimensionless black hole charge, and we require $0\leq e\leq1$. {The motion of a particle in the radial direction can be expressed} in a classical form \cite{Johnston}
\begin{eqnarray}
 \left(\rho^{2}\frac{d\hat{r}}{d\tau}\right)^{2}+V_{eff}=0,
\end{eqnarray}
with the effective potential $V_{eff}=-\hat{R}$. For a classical motion, we require $V_{eff}\leq0$ or $\hat{R}\geq0$. The turning point occurs at $V_{eff}=0$ or $\hat{R}=0$, at which the test particle has a zero radial velocity.
We plot the effective potential $V_{eff}$ against $\hat{r}$ in Fig.\ref{rhat1} by taking $e=0.2$, $\hat{L}/M\hat{E}=4$ and $Q/M^{2}\hat{E}^{2}$=3 as an example. For this case, the effective potential has two zero points at $\hat{r}_{1}=1.34M$ and $\hat{r}_{2}=3.32M$. The peak of the potential is at $\hat{r}_{p}=2.58M$. Note that the horizon locates at $\hat{r}_{h}=M$. For the photons coming from far region, they will stop at the point $\hat{r}_{2}$ and be reflected to infinity. And the photons emitted near the black hole horizon will be reflected at the point $\hat{r}_{1}$, and be absorbed by the black hole. While the region $\hat{r}\in(\hat{r}_{1}, \hat{r}_{2})$ is forbidden for such photons with such $\hat{L}$ and $Q$. There exists a special bound that the peak of the potential is shifted to the horizon, which will remove the region $(\hat{r}_{h}, \hat{r}_{p})$. Beyond this bound, the photons emitted near the horizon will freely travel to infinity without any turning point. Therefore, the photons will not have the chance to return back and take more energy from the black hole. This case is similar to the superradiant case used for a amplitude enhancement effect from the interaction between black holes and waves. In this paper, we will consider these trajectories near such bound in a KN black hole background. Moreover, such special bound can be obtained by solving $\hat{R}=\partial_{\hat{r}}\hat{R}=0$.

%%%%%%%%%%%%%%%%%%%%%%%%%%%%%%%%%%%%%%%%%%%%%%%%%%%%%%%%%%%%%%%%%%%%%
%%%%%%%%%%%%%%%%%%%%%%%%%%%%%%%%%%%%%%%%%%%%%%%%%%%%%%%%%%%%%%%%%%%%%%%%%

We define as did in Ref. \cite{Porfyriadis} a small dimensionless parameter
\begin{eqnarray}
 \lambda=1-\frac{\sqrt{1-e^{2}}}{1-e^{2}/2}\cdot\frac{\hat{L}}{2M\hat{E}}\ll 1.\label{lambda}
\end{eqnarray}
This result will recover the superradiant bound for the Kerr black hole with $e=0$. For $\lambda<0$, the photons emitted near the black hole horizon {will be reflected back by the potential, and could not approach  infinity. While when $\lambda>0$, these photons can escape the black hole. Thus $\lambda=0$ is a critical case. Here we expect to study the null geodesics with small $\lambda$. More interestingly, as did for the Kerr black hole \cite{Porfyriadis}, the geodesics (\ref{ttheta})-(\ref{ttheta2}) for the photons with small $\lambda$ emitted from the near-horizon region are analytically solvable to leading order in $\lambda$.}

Let us introduce the dimensionless Bardeen-Horowitz coordinates,
\begin{eqnarray}
 t=\frac{\hat{t}}{2M},\quad \phi=\hat{\phi}-\frac{\hat{t}}{2M},\quad
 r=\frac{\hat{r}-M}{M}.\label{newcoordinate}
\end{eqnarray}
After this coordinate transformation, the extremal black hole horizon is shifted to $r=0$. On the other hand, the energy $\hat{E}$ can be scaled out of the null geodesic with the shifted dimensionless Carter constant
\begin{eqnarray}
 q^{2}&=&\frac{3-4e^{2}}{1-e^{2}}-\frac{Q}{M^{2}\hat{E}^{2}}\nonumber\\
 &<& {\frac{4(1-\lambda+\lambda^{2})}{1-e^{2}}
  +\frac{\lambda^{2}e^{4}-2e^{2}(2-\lambda+2\lambda^{2})}{1-e^{2}},} \label{qq}
\end{eqnarray}
{where we have expressed the positivity of the kinetic energy in a local frame in the last inequality. In the equatorial plane, the photon trajectories has $q=\frac{(3-4e^{2})}{(1-e^{2})}$}. Here we are interested in the photon trajectories starting at $(t_{n}, r_{n}, \theta_{n}, \phi_{n})$ near the horizon and ending at $(t_{f}, r_{f}, \theta_{f}, \phi_{f})$, where a distant telescope locates. Then the geodesic equation can be expressed in terms of the new coordinates (\ref{newcoordinate}) as
\begin{eqnarray}
 \int_{r_{n}}^{r_{f}}\frac{dr}{\sqrt{R}}&=&\int_{\theta_{n}}^{\theta_{f}}\frac{d\theta}{\sqrt{\Theta}},\label{rtheta}\\
\phi_{f}\!-\!\phi_{n}&=& -\frac{1}{2}\int_{r_{n}}^{r_{f}}\frac{\Phi dr}{r^{2}\sqrt{R}}  \nonumber\\
      &+& {\frac{1}{2}\int_{\theta_{n}}^{\theta_{f}} \!
        \frac{2(2\!-\!e^{2})(1\!-\!\lambda)\!-\!(1\!-\!e^{2})^{3/2}\sin^{2}\theta}
             {\sqrt{1-e^{2}}\sqrt{\Theta}~\tan^{2}\theta}
         d\theta,} \nonumber\\ \label{phifphin}\\
 t_{f}\!-\!t_{n}&=&\frac{1}{2}\int_{r_{n}}^{r_{f}} \! \frac{T dr}{r^{2}\sqrt{R}}
             \!+\!\frac{1}{2}\int_{\theta_{n}}^{\theta_{f}}
             \frac{(1\!-\!e^{2})\cos^{2}\theta}{\sqrt{\Theta}}d\theta,\label{tftn}
\end{eqnarray}
where
\begin{eqnarray}
  R&=&r^{4}+4r^{3}+\frac{(1-e^{2})q^{2}+\lambda(2-e^{2})^{2}(2-\lambda)}{1-e^{2}}r^{2}\nonumber\\
       &+&4(2-e^{2})\lambda r+(2-e^{2})^{2}\lambda^{2},\nonumber\\
  \Theta&=&\bigg((3-q^{2})+e^{2}(q^{2}-4)+(1-e^{2})^{2}\cos^{2}\theta\nonumber\\
     &-&(2-e^{2})^{2}(1-\lambda)^{2}\cot^{2}\theta\bigg)/(1-e^{2}),\nonumber\\
  \Phi&=&r^{4}\!+\!4r^{3}
             \!+\!\frac{7\sqrt{1\!-\!e^{2}}\!+\!4\lambda\!-\!4\!-\!e^{2}(2\lambda\!+\!\sqrt{1\!-\!e^{2}}\!-\!2)}
                       {\sqrt{1-e^{2}}}r^{2}
       \nonumber\\
       &+&\bigg(4\left(1\!-\!\sqrt{1\!-\!e^{2}}\right)
         \!+\!4\lambda-2e^{2}(1\!+\!\lambda)\bigg)r\nonumber\\
         &+&(2\!-\!e^{2})\left[2(1\!-\!\sqrt{1\!-\!e^{2}})\!-\!e^{2}\right]\lambda,\nonumber\\
  T&=&r^{4}\!+\!4r^{3}\!+\!(7\!-\!e^{2})r^{2}\!+\!2(1\!+\!\lambda)(2\!-\!e^{2})r\!+\!(2\!-\!e^{2})^{2}\lambda.\nonumber
\end{eqnarray}
The integrals of (\ref{rtheta})-(\ref{tftn}) can be expressed with elliptic functions and are treated numerically ( for detail see Ref. \cite{Kraniotis}). However, in this paper, we will follow the technique of Ref. \cite{Porfyriadis} to analytically perform the integrals to the leading order in the small parameter $\lambda$ with the matched asymptotic expansions (MAE) method. The spacetime is divided into different regions \cite{Porfyriadis}:
\begin{eqnarray}
  &&\text{Near region:} \quad r\ll 1,\\
  &&\text{Far region:} \quad  r\gg \sqrt{\lambda},\\
  &&\text{Overlap region:} \quad \sqrt{\lambda}\ll r\ll 1.
\end{eqnarray}
Next, we will first solve these equations in the near and far regions, respectively, and then match them in the overlap region.

\section{The $r$-$\theta$ motion}
\label{rth}

In this section, we deal with the radial integral
\begin{eqnarray}
  I=\int_{r_{n}}^{r_{f}}\frac{dr}{\sqrt{R(r)}}.\label{rmotion}
\end{eqnarray}
This integral is also equal to that along the $\theta$ motion, $I=\int_{\theta_{n}}^{\theta_{f}}\frac{d\theta}{\sqrt{\Theta}}$. Taking the MAE approximation, we have
\begin{eqnarray}
  R(r)\approx r^{4}+4r^{3}+q^{2}r^{2}+4\lambda(2-e^{2}) r+(2-e^{2})^{2}\lambda^{2}.\;\;\;
\end{eqnarray}
In the near and far regions, one has
\begin{eqnarray}
  R_{n}(r)&=&q^{2}r^{2}+4\lambda(2-e^{2})r+(2-e^{2})^{2}\lambda^{2},\\
  R_{f}(r)&=&r^{2}(r^{2}+4r+q^{2}).
\end{eqnarray}
In the far region, the turning points in the radial motion {occur} at
\begin{eqnarray}
  r_{f\pm}=-2\pm\sqrt{4-q^{2}}.
\end{eqnarray}
This result is the same as that of the extremal Kerr black hole. In order to require the photon can penetrate the near region from the far one, there must be no turning point, and thus we take
\begin{eqnarray}
 q^{2}>0.\label{cond}
\end{eqnarray}
In the near region, the turning points are at
\begin{eqnarray}
  r_{n\pm}=\frac{(2-e^{2})\lambda}{q^{2}}\left(-2\pm\sqrt{4-q^{2}}\right).
\end{eqnarray}
{If $0<q^{2}<4$, there will be two turning points if and only if $\lambda$ is negative. And these geodesics originating from the horizon will turn back before they arrive the far region, which is because that the superradiant bound is exceed. However, if the geodesic starts outside the turning points, then it can reach the far region. On the other hand, when $q^{2}\geq4$, there will be no turning point outside the horizon for both negative or positive $\lambda$, and these photons can reach the far region. Therefore, when the condition (\ref{cond}) is satisfied, the geodesics with positive $\lambda$ can approach the horizon from a far region, while these with negative $\lambda$ will not get all the way to the black hole horizon.}

Now we can perform the radial integrals both in the near and far regions,
\begin{eqnarray}
  I_{n}(r)\!&=&\!\int^{r}\frac{dr'}{\sqrt{R_{n}(r')}}\nonumber\\
    &=&\!\frac{1}{q}\ln\left(q\sqrt{R_{n}(r)}+q^{2}r+2\lambda(2-e^{2})\right)+C_{n},\;\;\\
  I_{f}(r)\!&=&\!\int^{r}\frac{dr'}{\sqrt{R_{f}(r')}}\nonumber\\
    &=&\!-\frac{1}{q}\ln\frac{1}{r^{2}}\left(q\sqrt{R_{f}(r)}+q^{2}r+2r^{2}\right)+C_{f},
\end{eqnarray}
with $C_{n}$ and $C_{f}$ the integration constants. In the overlap region, the radial integrals are
\begin{eqnarray}
  I_{n}(r)&=&\frac{1}{q}\left(\ln r+\ln 2q^{2}+qC_{n}+\frac{2\lambda(2-e^{2})}{q^{2}r}+\cdots\right),\;\;\;\;\;\\
  I_{f}(r)&=&\frac{1}{q}\left(\ln r-\ln 2q^{2}+qC_{f}-\frac{2r}{q^{2}}+\cdots\right).
\end{eqnarray}
Taking the match in the overlap region, i.e., $I_{n}=I_{f}$, we have
\begin{eqnarray}
 C_{f}=C_{n}+\frac{2}{q}\ln 2q^{2}.
\end{eqnarray}
This condition is the same as that of the Kerr black hole given in Ref. \cite{Porfyriadis}. {Using it, we are allowed to calculate the integral (\ref{rmotion}) for the photon starting from the near region to the far one, which is given by
\begin{eqnarray}
  I&=&I_{f}(r_{f})-I_{n}(r_{n})\label{pp}\\
   &=&-\frac{1}{q}\ln\frac{(qD_{f}+q^{2}+2r_{f})(qD_{n}+q^{2}r_{n}+2\lambda(2-e^{2}))}
     {4q^{4}r_{f}},\nonumber
\end{eqnarray}}
where
\begin{eqnarray}
  D_{n}&=&\sqrt{R_{n}(r_{n})}\nonumber\\
        &=&\sqrt{q^{2}r_{n}^{2}+4\lambda(2-e^{2})r_{n}+(2-e^{2})^{2}\lambda^{2}},\\
  D_{f}&=&\frac{1}{r_{f}}\sqrt{R_{f}(r_{f})}
        =\sqrt{r_{f}^{2}+4r_{f}+q^{2}}.
\end{eqnarray}

\section{The $r$-$\phi$ and $r$-$t$ motions}
\label{rtrphi}

To leading order in $\lambda$ via MAE, we have
\begin{eqnarray}
 \Phi\!&=&\!r^{4}\!+\!4r^{3}
    \!+\!\left(7\!-\!\frac{4}{\sqrt{1\!-\!e^{2}}}
     \!+\!e^{2}\Big(\frac{2}{\sqrt{1\!-\!e^{2}}}\!-\!1\Big)\right)r^{2}\nonumber\\
    &&+\left(4\left(1-\sqrt{1-e^{2}}\right)+4\lambda-2e^{2}(1+\lambda)\right)r\nonumber\\
    &&+\lambda(2-e^{2})\left(2(1-\sqrt{1-e^{2}})-e^{2}\right),\\
 T\!&=&\!r^{4}\!+\!4r^{3}\!+\!(7\!-\!e^{2})r^{2}\!+\!2(2\!-\!e^{2})r\!+\!\lambda(2\!-\!e^{2})^{2}.\;\;\;\;
\end{eqnarray}
One can proceed to solve the radial integrals for the motion
\begin{eqnarray}
 I^{\phi}=\int_{r_{n}}^{r_{f}}\frac{\Phi}{r^{2}\sqrt{R(r)}},\quad
 I^{t}=\int_{r_{n}}^{r_{f}}\frac{T}{r^{2}\sqrt{R(r)}}.\label{tt0}
\end{eqnarray}
In the near region, one has
\begin{eqnarray}
 \Phi_{n}&=&\left(7-\frac{4}{\sqrt{1-e^{2}}}+e^{2}\left(\frac{2}{\sqrt{1-e^{2}}}-1\right)\right)r^{2}\nonumber\\
    &+&\left(4\left(1-\sqrt{1-e^{2}}\right)+4\lambda-2e^{2}(1+\lambda)\right)r\nonumber\\
    &+&\lambda(2-e^{2})\left(2\left(1-\sqrt{1-e^{2}}\right)-e^{2}\right),\\
 T_{n}&=&2(2-e^{2})r+(2-e^{2})^{2}\lambda.
\end{eqnarray}
In the far region, it gives
\begin{eqnarray}
 \Phi_{f}\!&=&\!r^{4}\!+\!4r^{3}
    \!+\!\left(7\!-\!\frac{4}{\sqrt{1\!-\!e^{2}}}\!+\!e^{2}\left(\frac{2}{\sqrt{1\!-\!e^{2}}}\!-\!1\right)\right)r^{2},\nonumber\\
 T_{f}&=&r^{4}\!+\!4r^{3}\!+\!(7\!-\!e^{2})r^{2}\!+\!2(2\!-\!e^{2})r.\nonumber
\end{eqnarray}
Now, all integrals can be performed,
\begin{eqnarray}
 I_{n}^{\phi}(r)&=&\int^{r}\frac{\Phi_{n}(r')}{r'^{2}\sqrt{R_{n}(r')}}dr'\nonumber\\
               &=&\frac{-4+7\sqrt{1-e^{2}}+e^{2}(2-\sqrt{1-e^{2}})}{q\sqrt{1-e^{2}}}\nonumber\\
                &\times&\ln\bigg(q\sqrt{R_{n}(r)}+q^{2}r+2(2-e^{2})\lambda\bigg)\nonumber\\
                &-&2\ln\frac{1}{r}\left(\sqrt{R_{n}(r)}+2r+2\lambda-e^{2}\lambda\right)\nonumber\\
                &-&\frac{(2-e^{2}-2\sqrt{1-e^{2}})\sqrt{R_{n}(r)}}{(2-e^{2})r\lambda}+C_{n}^{\phi},
                ~~~~~~~~~~~~~~~~~%\nonumber\\
\end{eqnarray}
\begin{eqnarray}
I_{f}^{\phi}(r)&=&\int^{r}\frac{\Phi_{n}(r')}{r'^{2}\sqrt{R_{f}(r')}}dr'\nonumber\\
           &=&-\frac{-4+7\sqrt{1-e^{2}}+e^{2}(2-\sqrt{1-e^{2}})}{q\sqrt{1-e^{2}}}\nonumber\\
                    &\times&\ln\frac{1}{r^{2}}\left(q\sqrt{R_{f}(r)}+q^{2}r+2r^{2}\right)\nonumber\\
                    &+&2\ln\frac{1}{r}\left(\sqrt{R_{f}(r)}+r^{2}+2r\right)\nonumber\\
                &+&\frac{1}{r}\sqrt{R_{f}(r)}
                   -2(2-e^{2}-2\sqrt{1-e^{2}})\nonumber\\
                &\times&\frac{q\sqrt{R_{f}(r)}\!+\!2r^{2}(2\ln r\!-\!\ln(q\sqrt{R_{f}(r)}\!+\!q^{2}r\!+\!2r^{2}))}
                             {q^{3}r^{2}}\nonumber\\
                &+&C_{f}^{\phi},%\nonumber\\
\end{eqnarray}
\begin{eqnarray}
 I_{n}^{t}(r)&=&\int^{r}\frac{T_{n}(r')}{r'^{2}\sqrt{R_{n}(r')}}dr'
               =-\frac{1}{\lambda r}\sqrt{R_{n}(r)}+C_{n}^{t}, ~~~~~~~ %\nonumber\\
 \end{eqnarray}
\begin{eqnarray}
 I_{f}^{t}(r)&=&\int^{r}\frac{T_{f}(r')}{r'^{2}\sqrt{R_{f}(r')}}dr'\nonumber\\
             &=&-\frac{7q^{2}-8+e^{2}(4-q^{2})}{q^{3}}\nonumber\\
               &\times&\ln\frac{1}{r^{2}}
               \left(q\sqrt{R_{f}(r)}+q^{2}r+2r^{2}\right)\nonumber\\
               &+&2\ln\frac{1}{r}\left(\sqrt{R_{f}(r)}+r^{2}+2r\right)\nonumber\\
               &+&\frac{q^{2}r-2(2-e^{2})}{q^{2}r^{2}}\sqrt{R_{f}(r)}
               +C_{f}^{t},%\nonumber
\end{eqnarray}
where $C_{n}^{\phi,t}$ and $C_{f}^{\phi,t}$ denote the integration constants. In the overlap region, we have
\begin{eqnarray}
 I_{n}^{\phi}(r)&=&\frac{1}{q}\bigg[\frac{7\sqrt{1-e^{2}}-4+e^{2}(2-\sqrt{1-e^{2}})}{\sqrt{1-e^{2}}}
                 (\ln r+\ln 2q^{2})\nonumber\\
                 &+&q(C_{n}^{\phi}-2\ln(q+2))\nonumber\\
                 &+&\frac{2\lambda(2-e^{2})((7-q^{2})\sqrt{1-e^{2}}+e^{2}(2-\sqrt{1-e^{2}})-4)}{q^{2}r\sqrt{1-e^{2}}}\nonumber\\
                 &-&\frac{2-e^{2}-2\sqrt{1-e^{2}}}{2(2-e^{2})}\nonumber\\
                 &\times&
                 \left(\frac{4(2\!-\!e^{2})}{r}\!+\!\frac{2q^{2}}{\lambda}
                 \!-\!\frac{(2\!-\!e^{2})^{2}(4\!-\!q^{2})\lambda}{q^{2}r^{2}}\right)
                 \!+\!\cdots\bigg],%\nonumber\\
 \end{eqnarray}
\begin{eqnarray}
 I_{f}^{\phi}(r)&=&\frac{1}{q}\bigg[\frac{7\sqrt{1-e^{2}}-4+e^{2}(2-\sqrt{1-e^{2}})}{\sqrt{1-e^{2}}}
                 (\ln r-\ln 2q^{2})\nonumber\\
                 &+&q(C_{f}^{\phi}+2\ln(q+2))+q^{2}\nonumber\\
                 &+&\frac{24(\sqrt{1-e^{2}}-1)+10q^{2}}{q^{4}\sqrt{1-e^{2}}}r\nonumber\\
                 &-&\frac{(4-q^{2})(4q^{2}\sqrt{1-e^{2}}-3e^{2}(2-\sqrt{1-e^{2}}))}{q^{4}\sqrt{1-e^{2}}}r\nonumber\\
                 &-&2(2\!-\!e^{2}\!-\!2\sqrt{1\!-\!e^{2}})\frac{q^{2}\!+\!2r\!+\!2r\ln(\frac{r}{2q^{2}})}{q^{2}r}
                        \!+\!\cdots\bigg],%\nonumber\\
 \end{eqnarray}
\begin{eqnarray}
 I_{n}^{t}(r)=\!-\!\frac{1}{q}\!\left[\frac{q^{2}}{\lambda}\!-\!qC_{n}^{t}\!+\!\frac{4-2e^{2}}{r}
              \!+\!\frac{\lambda(2\!-\!e^{2})^{2}(q^{2}\!-\!4)}{2q^{2}r^{2}}\!+\!\cdots\!\right],\nonumber\\
  \end{eqnarray}
\begin{eqnarray}
 I_{f}^{t}(r)&=&-\frac{1}{q}\bigg[\frac{2(2-e^{2})}{r}
               -\frac{7q^{2}-8+e^{2}(4-q^{2})}{q^{2}}\ln r \nonumber\\
               &+&\frac{(7q^{2}-8+e^{2}(4-q^{2}))\ln 2q^{2}-q^{4}+8-4e^{2}}{q^{2}}\nonumber\\
              &-&q(C_{f}^{t}+2\ln (q+2))\nonumber\\
              &-&\frac{4(q^{4}-4q^{2}+6)-3e^{2}(4-q^{2})}{q^{4}}r+...\bigg]. ~~~~~~~~~~~~~~~~~%\nonumber
\end{eqnarray}
Matching $I_{n}^{\phi,t}=I_{f}^{\phi,t}$, the relations between the integration constants are
\begin{eqnarray}
 C_{f}^{\phi}&=&C_{n}^{\phi}+2\frac{4(1-\sqrt{1-e^{2}})+(7\sqrt{1-e^{2}}-4)q^{2}}{\sqrt{1-e^{2}}q^{3}}\ln2q^{2}\nonumber\\
                  &-&2\frac{e^{2}(2-q^{2})(2-\sqrt{1-e^{2}})}{\sqrt{1-e^{2}}q^{3}}\ln2q^{2}\nonumber\\
        &-&4\ln(2+q)-\frac{2(1+\lambda-\sqrt{1-e^{2}})-e^{2}(1+\lambda)}{(2-e^{2})\lambda}q \nonumber\\
        &+&\frac{4(2-e^{2}-2\sqrt{1-e^{2}})}{q^{3}},\\%\nonumber\\
 C_{f}^{t}&=&C_{n}^{t}-\frac{q}{\lambda}-2\ln(q+2)\nonumber\\
 &-&\frac{q^{4}+4e^{2}-8-(7q^{2}-8+e^{2}(4-q^{2}))\ln2q^{2}}{q^{3}}\nonumber\\
        &\approx& C_{n}^{t}-\frac{q}{\lambda}.%\nonumber
\end{eqnarray}
{Then the integrals in Eq. (\ref{tt0}) are}
\begin{eqnarray}
 I^{\phi}&=&-\frac{-4+7\sqrt{1-e^{2}}+e^{2}(2-\sqrt{1-e^{2}})}{q\sqrt{1-e^{2}}}\nonumber\\
                 &\times&\ln\frac{1}{r_{f}}(qD_{n}+q^{2}r_{n}+2(2-e^{2})\lambda)(qD_{f}+2r_{f}+q^{2})\nonumber\\
                 &+&2\ln\frac{1}{r_{n}}(D_{n}+2r_{n}+(2-e^{2})\lambda)(D_{f}+r_{f}+2)
                 \nonumber\\&&+D_{f}
                 +\frac{4(1-\sqrt{1-e^{2}})+(7\sqrt{1-e^{2}}-4)q^{2}}{\sqrt{1-e^{2}}}\frac{\ln 4q^{4}}{q^{3}}\nonumber\\
                 &-&\frac{e^{2}(2-\sqrt{1-e^{2}})(q^{2}-2)}{\sqrt{1-e^{2}}}\frac{\ln 4q^{4}}{q^{3}}
                 -4\ln(q+2)\nonumber\\
                 &-&\frac{2(1+\lambda-\sqrt{1-e^{2}})-e^{2}(1+\lambda)}{\lambda(2-e^{2})}q\nonumber\\
                 &+&\frac{2-e^{2}-2\sqrt{1-e^{2}}}{q^{3}}
                 \bigg(4-\frac{2qD_{f}}{r_{f}}+\frac{q^{3}D_{n}}{r_{n}\lambda(2-e^{2})}\nonumber\\
                 &+&4\ln\frac{1}{r_{f}}(qD_{f}+2r_{f}+q^{2})\bigg),\label{fphi}
\end{eqnarray}
\begin{eqnarray}
 I^{t}&=&2\ln(D_{f}+r_{f}+2)\nonumber\\
                &-&\frac{7q^{2}-8+e^{2}(4-q^{2})}{q^{3}}\ln\frac{1}{r_{f}}(qD_{f}+2r_{f}+q^{2})\nonumber\\
                &+&\frac{q^{2}r_{f}-2(2-e^{2})}{q^{2}r_{f}}D_{f}
                +\frac{1}{\lambda r_{n}}D_{n}-\frac{q}{\lambda}.\label{ft}
\end{eqnarray}
Finally, we can get the integrals in Eqs. (\ref{phifphin}) and (\ref{tftn}) when the $\theta$ integrals drop. In particular, for the case that the photons are limited in the equatorial plane $(\theta=\pi/2)$, the shifts in  the azimuthal angle $\phi$ and time $t$ can be calculated as
\begin{eqnarray}
 \phi_{f}-\phi_{n}&=&-\frac{1}{2}I^{\phi}\big|_{q=\sqrt{\frac{3-4e^{2}}{1-e^{2}}}},\label{fphi2}\\
 t_{f}-t_{n}&=&\frac{1}{2}I^{t}\big|_{q=\sqrt{\frac{3-4e^{2}}{1-e^{2}}}}.\label{ft2}
\end{eqnarray}
For the case that the photons are emitted from a distant source, pass by near a black hole, and finally are observed by a distant observer, their deflection angle and time delay can be measured with Eqs. (\ref{fphi2}) and (\ref{ft2}) {if the photon trajectories are limited in the equatorial plane}. Correspondingly, we assume one endpoint is near the photon sphere of the black hole and other one is located at infinity, then we have the deflection angle and time delay in the strong gravitational limit for the photons
\begin{eqnarray}
 \Delta \phi&=&2(\phi_{f}-\phi_{n})-\pi,\\
 \Delta t&=&2(t_{f}-t_{n}),
\end{eqnarray}
where the factor 2 comes from the symmetry analysis of the geodesics.

\section{Summary}

In this paper, we considered a special class of null geodesics in an extremal KN black hole background. Adopting the matched asymptotic expansion method, we performed the integrals for the null geodesic and derived the algebraic equations, see Eqs. (\ref{pp}), (\ref{fphi}), and (\ref{ft}). These equations are parameterized by $(\lambda, q)$ given in (\ref{lambda}) and (\ref{qq}). They relate two endpoints, one is near the horizon $(t_{n}, r_{n}, \phi_{n})$ and another locates at a far region from the black hole $(t_{f}, r_{f}, \phi_{f})$. The photons are free of the polar angle when they move from the near region to the far one. A stronger constraint is that the photon trajectory is limited in the equatorial plane $\theta=\pi/2$, which is also a simple requirement when considering the strong black hole lensing.

As pointed out in Refs. \cite{Porfyriadis,Zahrani}, the parameter $\Theta$ must be positive. For the extremal Kerr black hole, $\theta$ must be in the region $\theta \in (47^{\circ}, 133^{\circ})$. While for the extremal KN black hole, such region is modified by the black hole charge. For example, the condition $3+(1-e^{2})\cos^{2}\theta-4\cot^{2}\theta\geq0$ must be held, which implies that $\theta$ should lie between $\theta_{0}=\arccos\sqrt{\frac{\sqrt{e^{4}+48}-6-e^{2}}{2(1-e^{2})}}$ and $\pi-\theta_{0}$. When the black hole charge vanishes, it recovers the region for the Kerr black hole. On the other hand, when the black hole charge approaches its maximum $e=1$, the region will approximately be $(49^{\circ}, 131^{\circ})$. A detailed analysis suggests that this region shrinks with the increase of the black hole charge.

Since our analysis of the null geodesics is near the critical point that the photons can escape from the black hole, many astronomical phenomena are related with it, such as the black hole lensing and shadow. So using the analytical results obtained in this paper, one can consider these phenomena in an analytical way for the extremal black hole.

\section*{Acknowledgements}

This work was supported by the National Natural Science Foundation of China (Grants No. 11675064, No. 11522541, and No. 11375075), and the Fundamental Research Funds for the Central Universities (No. lzujbky-2016-121 and lzujbky-2016-k04).


\begin{thebibliography}{99}

\bibitem{LIGO}
 The Laser Interferometer Gravitational-Wave Observatory (LIGO),
http://www.ligo.org.

\bibitem{EHT}
 The Event Horizon Telescope (EHT), http://www.eventhorizontelescope.org.

\bibitem{ATHENA}
The Advanced Telescope for High ENergy Astrophysics (ATHENA), http://www.the-
athena-x-ray-observatory.eu.

\bibitem{SKA}
 The Square Kilometre Array (SKA), http://www.skatelescope.org.

\bibitem{eLISA}
The Evolved Laser Interferometer Space Antenna (eLISA), http://www.elisascience.org.

\bibitem{Eddington}
 A. S. Eddington, {\em Internal Constitution of the Stars} (Cambridge University Press, 1926); N. K. Glendenning, {\em Compact Stars} (AA Library, Springer-Verlag New York Inc., 2000), p. 82.

\bibitem{Ghezzi}
 C. R. Ghezzi,
  {\em Relativistic structure, stability, and gravitational collapse of charged neutron stars},
    Phys. Rev. \textbf{D 72}, 104017 (2005),
     [arXiv:gr-qc/0510106].

\bibitem{Letelier}
 C. R. Ghezzi and P. S. Letelier,
   {\em Numeric simulation of relativistic stellar core collapse and the formation of Reissner-Nordstrom black holes},
     Phys. Rev. \textbf{D 75}, 024020 (2007),
      [arXiv:astro-ph/0503629].

\bibitem{Cuesta}
 H. J. M. Cuesta, A. Penna-Firme, and A. Perez-Lorenzana,
  {\em Charge asymmetry in the brane world and the formation of charged black holes}, Phys. Rev. \textbf{D 67}, 087702 (2003),
   [arXiv:hep-ph/0203010].


\bibitem{Damour}
 T. Damour, R. S. Hanni, R. Ruffini, and J. R. Wilson,
 {\em Regions of magnetic support of a plasma around a black-hole},
   Phys. Rev. \textbf{D 17}, 1518 (1978).

\bibitem{Ruffini}
 R. Ruffini, G. Vereshchagin, and S.-S. Xue,
  {\em Electron-positron pairs in physics and astrophysics: From heavy nuclei to black holes},
    Phys. Rept. \textbf{487}, 1 (2010),
    [arXiv:0910.0974[astro-ph.HE]].

\bibitem{Johnston}
 M. Johnston and R. Ruffini,
  {\em Generalized Wilkins effect and selected orbits in a Kerr-Newman geometry}, Phys. Rev. \textbf{D 10}, 2324 (1974).

\bibitem{Young}
  P. J. Young,
   {\em Capture of particles from plunge orbits by a black hole},
    Phys. Rev. \textbf{D 14}, 3281 (1976).



\bibitem{Deriglazov}
 A. A. Deriglazov and W. G. Ramirez,
  {\em World-line geometry probed by fast spinning particle},
   Mod. Phys. Lett. \textbf{A 30}, 1550101 (2015),
   [arXiv:1409.4756[physics.gen-ph]].

\bibitem{Bicak}
 V. Balek, J. Bicak, and Z. Stuchlik,
 {\em The motion of charged particles in the field of rotating charged black holes and naked singularities-II. the motion in the equatorial plane},
  Bull. Astron. Inst. Czechosl. \textbf{40}, 133 (1989).

\bibitem{Kovar}
 J. Kovar, Z. Stuchlik, and V. Karas,
 {\em Off-equatorial orbits in strong gravitational fields near compact objects}, Class. Quant. Grav. \textbf{25}, 095011 (2008).

\bibitem{Calvani}
 M. Calvani and R. Turolla,
  {\em Complete description of photon trajectories in the Kerr-Newman space-time},
    J. Phys. A: Math. Gen. \textbf{14}, 1931 (1981).

\bibitem{Hackmann}
 E. Hackmann and H. Xu,
  {\em Charged particle motion in Kerr-Newman space-times},
   Phys. Rev. \textbf{D 87}, 124030 (2013),
    [arXiv:1304.2142[gr-qc]].

\bibitem{Kraniotis}
 G. V. Kraniotis,
  {\em Gravitational lensing and frame dragging of light in the Kerr-Newman and the Kerr-Newman-(anti) de Sitter black hole spacetimes},
   Gen. Rel. Grav. \textbf{46}, 1818 (2014),
    [arXiv:1401.7118[gr-qc]].

\bibitem{Hartman}
 T. Hartman, K. Murata, T. Nishioka, and A. Strominger,
 {\em CFT Duals for Extreme Black Holes},
  JHEP \textbf{0904}, 019 (2009),
       [arXiv:0811.4393[hep-th]].

\bibitem{Song}
   T. Hartman, W. Song, and A. Strominger,
   {\em Holographic Derivation of Kerr-Newman Scattering Amplitudes for General
Charge and Spin},
 JHEP \textbf{1003}, 118 (2010),
   [arXiv:0908.3909[hep-th]].

\bibitem{Rasmussen}
 J. Rasmussen,
 {\em On the CFT duals for near-extremal black holes},
   Mod. Phys. Lett. \textbf{A 26}, 1601 (2011),
    [arXiv:1005.2255[hep-th]].

\bibitem{Chen}
 C.-M. Chen, Y.-M. Huang, J.-R. Sun, M.-F. Wu, and S.-J. Zou,
 {\em Twofold Hidden Conformal Symmetries of the Kerr-Newman Black Hole},
   Phys. Rev. \textbf{D 82}, 066004 (2010),
     [arXiv:1006.4097[hep-th]].

\bibitem{ChenChen}
 B. Chen, C.-M. Chen, and B. Ning,
 {\em Holographic Q-picture of Kerr-Newman-AdS-dS Black Hole},
   Nucl. Phys. \textbf{B 853}, 196 (2011),
    [arXiv:1010.1379[hep-th]].




\bibitem{Sun}
 C.-M. Chen and J.-R. Sun,
  {\em The Kerr-Newman/CFTs Correspondence},
    Int. J. Mod. Phys. Conf. Ser. \textbf{07}, 227 (2012),
     [arXiv:1201.4040[hep-th]].

\bibitem{Bai}
 N. Bai, Y.-H. Gao, X.-B. Xu,
  {\em On neutral scalar radiation by a massive orbiting star in extremal Kerr-Newman black hole},
    Fortsch. Phys. \textbf{63}, 323 (2015),
     [arXiv:1407.0089[hep-th]].

\bibitem{Guica}
 M. Guica, T. Hartman, W. Song, and A. Strominger,
   {\em The Kerr/CFT Correspondence},
     Phys. Rev. \textbf{D 80}, 124008 (2009),
      [arXiv:0809.4266[hep-th]].

\bibitem{Bredberg}
 I. Bredberg, T. Hartman, W. Song, and A. Strominger,
   {\em Black Hole Superradiance From Kerr/CFT},
     JHEP \textbf{1004}, 019 (2010),
       [arXiv:0907.3477[hep-th]].

\bibitem{Compere}
 G. Compere,
   {\em The Kerr/CFT correspondence and its extensions: a comprehensive review}, Living Rev. Rel. \textbf{15}, 11 (2012),
     [arXiv:1203.3561[hep-th]].

\bibitem{Chiang}
  C.-M. Chen, S. P. Kim, J.-R. Sun, and F.-Y. Tang,
    {\em Pair Production in Near Extremal Kerr-Newman Black Holes},
     [arXiv:1607.02610[hep-th]].

\bibitem{Porfyriadis}
  A. P. Porfyriadis, Y. Shi, and A. Strominger,
   {\em Photon Emission Near Extreme Kerr Black Holes},
     [arXiv:1607.06028[gr-qc]].


\bibitem{Carter}
 B. Carter,
  {\em Global structure of the Kerr family of gravitational fields},
  Phys. Rev. \textbf{174}, 1559 (1968).


\bibitem{Zahrani}
 A. M. Al Zahrani, V. P. Frolov, and A. A. Shoom,
  {\em Particle Dynamics in Weakly Charged Extreme Kerr Throat},
  Int. J. Mod. Phys. \textbf{D 20}, 649 (2011), [arXiv:1010.1570[gr-qc]].



\end{thebibliography}
\end{document}